\PassOptionsToPackage{pdfpagelabels=false}{hyperref}
\documentclass[useAMS,usenatbib]{mnras}
\pdfoutput=1 
\usepackage{graphicx}
\usepackage{amsmath,amssymb,amstext}
\usepackage[T1]{fontenc}
\usepackage{ae,aecompl}
\usepackage[utf8]{inputenc}
\usepackage{newtxtext,newtxmath}
\usepackage[figure,figure*]{hypcap}
\usepackage[dvipsnames]{xcolor} 
\usepackage{bm}
\usepackage{multicol}
\usepackage{xparse}
\usepackage{url}
\usepackage{cuted}
\usepackage{xcolor}

\newcommand{\rhox}{\rho_\chi}
\newcommand{\vesc}{v_{\text{e}}}
\newcommand{\vp}{v^\prime}
\newcommand{\mpr}{m^\prime}
\newcommand{\xp}{x^\prime}
\newcommand{\SA}{S_{\text{A}}}
\newcommand{\SB}{S_{\text{B}}}
\newcommand{\SC}{S_{\text{C}}}
\newcommand{\G}{\text{G}}
\newcommand{\mx}{m_\chi}
\newcommand{\rx}{r_\chi}
\newcommand{\Nx}{N_\chi}
\newcommand{\MDM}{M_{\mathrm{DM}}}
\newcommand{\Rvir}{R_{\mathrm{vir}}}
\newcommand{\Msun}{\textrm{M}_\odot}
\newcommand{\kpc}{\textrm{kpc}}
\newcommand{\pc}{\textrm{pc}}
\newcommand{\gev}{\textrm{GeV}}
\newcommand{\Gev}{\textrm{GeV}}
\newcommand{\km}{\textrm{km}}
\newcommand{\cm}{\textrm{cm}}
\newcommand{\s}{\textrm{s}}

\begin{document}

\title[DM Escape from GCs]{On the Ejection of Dark Matter from Globular Clusters}

\author[Hurst \& Zentner]{%
Travis J. Hurst,$^{1}$\thanks{E-mail: travis.hurst@csupueblo.edu}
\& Andrew R. Zentner,$^{2}$
\vspace*{12pt}
\\
$^{1}$Department of Mathematics and Physics, Colorado State University - Pueblo, Pueblo, CO 81001\\
$^{2}$Department of Physics and Astronomy and The Pittsburgh Particle physics, Astrophysics, and Cosmology Center (Pitt-PACC), \\ \ \ \ University of Pittsburgh, Pittsburgh, PA 15260, USA
}
\maketitle

\begin{abstract}
We investigate analytically whether in a close encounter with a star, a Dark Matter particle can be accelerated above the escape speed of a Globular Cluster and be ejected.  We find that this mechanism is not sufficient to eject a massive, extended Dark Matter halo by the present time. Combined with observations of isolated Globular Clusters that may not have had their halos tidally stripped, these results cast doubt on the scenario in which Globular Clusters formed in Dark Matter halos.
\end{abstract}

\maketitle

\section{INTRODUCTION}
\label{section:intro3}

In the $\Lambda$CDM paradigm, Dark Matter (DM) is the first matter constituent to collapse, forming DM halos which serve as the seeds for galaxy formation. Progressively larger structures are built through the mergers of halos. This hierarchical structure formation is predicted by theory and is seen in $N$-body simulations, as well as observed in the structure of galaxies and galaxy clusters. 

One seeming exception to this scenario are Globular Clusters (GC). \cite{Peebles1984} was the first to propose that GCs form in extended DM halos. However, observations of many GCs reveal thin tidal tails that $N$-body simulations predict should not form if they possess halos \cite{Grillmair1995,Odenkirchen2003,Moore1996}. Moreover, recent studies of several GCs indicate that the ratio of the mass in DM to stars in several GCs is less than unity ($\MDM/M_* \lesssim 1$) \cite{Shin, Conroy,Ibata2013} and is potentially $\lesssim 10^{-2}$ if the DM is a low mass ($\mx \sim 10\, \gev$) weakly interacting particle \cite{HurstWDGC}. 

It is now generally thought GCs formed in gas compressed by shocks \cite{Gunn1980,Harris1994}. However, the formation scenarios of GCs remain controversial in part because of the complex abundance patterns measured in stars. These observations indicate that GCs must have been much more massive in the past in order to retain significant amounts of heavy elements that would have been ejected by supernovae \cite{Gratton,Gratton2012,Con&Sperg}. As pointed out in \cite{Conroy} this formation scenario is further complicated by the existence of nuclear star clusters, which demonstrates that at least some GC-like systems form in DM halos (e.g. \cite{Taylor2015,Boker2004,Walcher2005,Walcher2006}).

As the resolution of $N$-body simulations has improved it has become possible to study small stellar populations such as GCs \cite{Phipps2019}. Recently several groups have made progress on identifying GCs in $N$-body simulations \cite{Ishiyama2013, Rieder2013, Kim2018, Carlberg2018, Reina2019, Li2019, Ma2019, Phipps2019}. In particular \cite{Phipps2019} identify infant GCs at $z \geq 6$ within the cosmological simulations from the First Billion Years (FiBY) project. These authors suggest that GCs formed from high baryon fraction systems embedded within the peaks of their Galaxy's host halos. Therefore, these systems may possess a small number of energetically bound DM particles, yet they still formed in high DM density environments.

Though they seemingly do not possess DM halos today, GCs could have had them in the past and subsequently lost their DM. One mechanism invoked for the removal of the halo is tidal stripping by the galaxy \cite{Bromm2002,Mashchenko2005}. While the majority of the Galactic Globular Clusters (GGC) orbit within strong tidal fields, there does exist a population of isolated GCs with galactocentric distances $r_{\mathrm{gc}} > 70\, \kpc$ that should not have lost their halos through tidal interactions. Two such GCs are NGC 2419 ($r_{\mathrm{gc}} = 89.9\, \kpc$) and MGC1, which at $\sim 200\, \kpc$ from M31 is the most isolated cluster in the local group \cite {Harris, Conroy,Mackey2010}. Observations of both these clusters indicate that $\MDM/M_* \lesssim 1$ \cite{Conroy,Ibata2013}. Note that several clusters that meet this definition of isolated actually do show evidence for tidal structures, so it can not be ruled out that clusters with $r_{\mathrm{gc}} > 70\, \kpc$ have had their hypothetical halos tidally stripped \cite{Sohn2003,Myeong2017,Sollima2011}. In particular, Palomar 4 shows evidence for tidal structures despite a galactocentric distance $r_{\mathrm{gc}} = 111.4\, \mathrm{kpc}$ \cite{Sohn2003,Harris}.

There are several channels through which GCs could eject DM halos, e.g. DM decay \cite{Peter2010} and stellar feedback \cite{Davis2014}. In this paper we investigate the following mechanism: In a close encounter with a star, a DM particle can be accelerated above the escape speed of the GC and be ejected. In principle, the DM halo can also \emph{evaporate} if a DM particle slowly builds up speed through multiple interactions with stars. However, this mechanism is not efficient in GCs because particles with velocities near the escape speed spend most of their time near the outskirts of the GC and therefore, rarely experience an encounter with a star \cite{HenonB}. Of course, a star does not interact with a single DM particle, but with many, so the ejection rate is closely related to the dynamical friction timescale. In \cite{Baumgardt2008} it was found using $N$-body simulations that the inner portions of the halos of GCs should be scattered to beyond the half-mass radius in significantly less than a Hubble time. Note that this provides more opportunity for the halo to be tidally stripped. 

In this paper we will investigate the escape rate of DM particles from a spherically symmetric stellar system in order to ascertain the viability of the ejection scenario. As the interaction is gravitational, we shall not trouble ourselves with the details of the DM particle. The only assumption we make of the DM particle is that its mass is significantly less than the mass of a typical star.

The remainder of the paper is organized as follows: in \S\ref{section:methods3} we outline the calculation of the escape rate of DM particles from an isolated, spherical stellar system. In \S\ref{section:results3} we present our results, and in \S\ref{section:conclusions3} we discuss our conclusions.

\section{METHODS}
\label{section:methods3}

Our calculation will follow the approach of a pair of classic papers by H\'{e}non (\cite{HenonA,HenonB}). We begin with the assumption that the DM and stellar distributions are spherically symmetric and that the particle velocities are isotropic. Then, the number of DM particles in a phase space volume element $d^3rd^3v$ is
%
\begin{equation}
(4\pi)^2r^2v^2f(r,v)d rd v,
\end{equation}
where $f(r,v)$ is the DM distribution function. Similarly, if the stellar distribution function is $g(r,\vp,\mpr$) then the number of stars in the volume element $d^3rd^3\vp d \mpr$ is
%
\begin{equation}
(4\pi)^2r^2{\vp}^2g(r,\vp,\mpr)d rd \vp d \mpr.
\end{equation}

Consider a DM particle of mass $\mx$ and coordinates $(r,v)$. According to \cite{HenonA} the probability that a particle will experience an encounter that takes it from a velocity $\vec{v} \rightarrow \vec{v} + \vec{e}$ is
%
\begin{equation}
P = 8\pi \G^2d t\frac{d^3e}{e^5}\int^\infty_0 {\mpr}^2\,d \mpr\int^\infty_{\vp_0}g(r,\vp,\mpr)\vp\,d \vp, 
\end{equation}
where G is Newton's constant and the lower limit $\vp_0 = \frac{1}{e}|\vec{v}\cdot\vec{e} + \frac{\mx+\mpr}{2\mpr}e^2|$ can be thought of as a statement of conservation of momentum. 

As stated in \S\ref{section:intro3}, the one assumption of the DM particle we make is that $\mx \ll \mpr$ so 
%
\begin{equation}
\label{v0p1}
\begin{split}
\vp_0 &= \frac{1}{e}\Bigg|\vec{v}\cdot\vec{e} + \frac{e^2}{2}\Bigg|, \\
&= \bigg|v\cos\delta + \frac{e}{2}\bigg|.
\end{split}
\end{equation}
The particle will escape if 
%
\begin{equation}
\big|\vec{v}+\vec{e}\big| \ge \vesc(r),
\label{geve}
\end{equation}
where $\vesc(r)$ is the local escape velocity. In the remainder of the paper we will denote the local escape velocity simply as $\vesc$. Using the notation of \cite{HenonB}, let $e, \delta, \varphi$ be a set of spherical coordinates for the kick velocity $\vec{e}$. Then from Equation~(\ref{geve}), the condition for escape is
\begin{equation}
\label{geve2}
v^2 + e^2 + 2ve\cos\delta \ge \vesc^2.
\end{equation}
Then we can write the probability that the DM particle will escape in a time d$t$ as:
%
\begin{equation}
    \begin{split}
Q = 8\pi\G^2d t\int^\infty_0 & {\mpr}^2\,d \mpr\int^\infty_{\vp_0} g(r,\vp,\mpr)\vp\,d \vp \\ 
    & \times \int^{2\pi}_0{}\,d \varphi\int{\sin\delta}\,d \delta\int{e^{-3}}\,d e.
\end{split}
\end{equation}
For a bound DM particle it must be the case that $v < \vesc$, then from (\ref{geve2})
%
\begin{equation}
\vesc^2 \le v^2 + e^2 + 2ve\cos\delta \le \vesc^2 + e^2 + 2ve\cos\delta,
\end{equation}
therefore,
%
\begin{equation}
v\cos\delta \ge \frac{-e}{2}.
\end{equation}
Hence, we can drop the absolute value in (\ref{v0p1}). Now,
%
\begin{equation}
    \begin{split}
Q = 16\pi^2\G^2d t\int^\infty_0 & {\mpr}^2\,d \mpr\int^\infty_{\vp_0}{g(r,\vp,\mpr)\vp}\,d \vp \\ 
    & \times \int{e^{-3}}\,d e\int{}\,d \cos\delta,
\end{split}
\end{equation}

where integration should satisfy:
%
\begin{equation}
-1 \le \cos\delta \le 1,
\label{lim1}
\end{equation}
\begin{equation}
0 \le e,
\label{lim2}
\end{equation}
\begin{equation}
v^2 + e^2 + 2ve\cos\delta \ge \vesc^2,
\label{lim3}
\end{equation}
\begin{equation}
v\cos\delta + \frac{e}{2} \le \vp < \vesc.
\label{lim4}
\end{equation}
To find the escape rate, we now integrate over the position and velocity of the DM particle. Let $\Nx$ be the number of DM particles in the cluster, 
%
\begin{equation}
\Nx = \int^\infty_0{4\pi r^2}\,d r\int^{\vesc}_0{4\pi v^2f(r,v)}\,d v\int^\infty_0{\Nx(m)}\,d m,
\label{Nchi}
\end{equation}
with $f(r,v)$ normalized to 1 and $\Nx(m) = \Nx\delta(m-\mx)$ assuming the halo is composed of a single DM constituent. Then the specific escape rate is
%
\begin{equation}
\label{esc1}
\begin{split}
\bigg|\frac{1}{\Nx} & \frac{\partial \Nx}{\partial t}\bigg|  = \int^\infty_0{4\pi r^2}\,d r\int^{\vesc}_0{4\pi v^2\frac{Q}{d t}f(r,v)}\,d v, \\
& = 256\pi^4\G^2\int^\infty_0{r^2}\,d r\int^{\vesc}_0{v^2f(r,v)}\,d v \int^\infty_0{\mpr}^2\,d \mpr \\
& \ \ \ \ \ \ \ \ \ \ \ \times \int^\infty_{\vp_0}{g(r,\vp,\mpr)\vp}\,d \vp\int{e^{-3}}\,d e\int{}\,d\cos\delta,
\end{split}
\end{equation}
with the limits in Equations(\ref{lim1})-(\ref{lim4}) satisfied and where we have taken the magnitude since $\frac{\partial \Nx}{\partial t}$ is negative. If the magnitude of the specific escape rate is greater than $\tau^{-1}$ with $\tau$ the age of the Universe, then a typical DM particle will have been ejected from the halo.  It is therefore likely that the GC would have dissipated its halo by the present time {\em via} this mechanism. We normalized Equation~(\ref{Nchi}) to $\Nx$ rather than 1 to make this point explicit. 

This expression looks quite intractable, but the integrals in $e$ and $\delta$ can in fact be calculated analytically. Keeping with the notation of ~\cite{HenonB} let
\begin{equation}
S = \int{e^{-3}}\,d e\int{}\,d \cos\delta,
\end{equation}
and let $C = \cos\delta$. From (\ref{lim3})
\begin{equation}
C \ge \frac{\vesc^2-v^2-e^2}{2ve} = C_1,
\end{equation}
from (\ref{lim4})
\begin{equation}
C \le \frac{\vp - \frac{e}{2}}{v} = C_2,
\end{equation}
and from (\ref{lim1})
\begin{equation}
C_3 = -1 \le C \le 1 = C_4.
\end{equation}
In order for $S$ to be non-zero we must have that $C_1 < C_4, C_1 < C_2, C_3 < C_4$, and $C_3 < C_2$. Now $C_3 < C_4$ trivially. $C_1 < C_4$ requires that,
%
\begin{equation}
e > \vesc - v = e_1,
\label{e1}
\end{equation}
which is stronger than (\ref{lim2}). $C_1 < C_2$ requires that,
\begin{equation}
e > \frac{\vesc^2 - v^2}{2\vp} = e_2,
\label{e2}
\end{equation}
which is again stronger than (\ref{lim2}). And $C_3 < C_2$ requires that,
\begin{equation}
e < 2(\vp + v) = e_3,
\label{e3}
\end{equation}
which further restricts (\ref{lim2}). $C_3$ will be the lower limit of the $d C$ integral when $C_1 < C_3$ or when 
\begin{equation}
e > v + \vesc = e_4,
\end{equation}
and $C_2$ will be the upper limit when $C_2 < C_4$ or when 
\begin{equation}
e > 2(\vp - v) = e_5. 
\end{equation}
Thus, in order to determine the limits of the integrals in $S$, we must consider the order of $e_1, e_2, e_3, e_4$, and $e_5$. Elementary calculations show that 
%
\begin{equation}
\begin{split}
\vp \ge \frac{1}{2}(\vesc - 3v) = \vp_1 &\Rightarrow e_1 \le e_3,\\
\vp \ge \frac{1}{2}(\vesc - v) = \vp_2 &\Rightarrow e_2 \le e_3, e_2 \le e_4, e_4 \le e_3,\\
\vp \ge \frac{1}{2}(\vesc + v) = \vp_3 &\Rightarrow e_2 \le e_1, e_1 \le e_5, e_2 \le e_5,\\
\vp \ge \frac{1}{2}(\vesc + 3v) = \vp_4 &\Rightarrow e_4 \le e_5,\\
\end{split}
\end{equation}
and it is always true that $e_1 \le e_4$ and $e_5 \le e_3$. These relations divide the $v$-$ \vp$ plane into 5 regions A, B, C, D, and E. In region A,
%
\begin{equation}
e_5 \le e_1 \le e_2 \le e_4 \le e_3. 
\end{equation} 
Thus in region A we have,
%
\begin{equation}
\begin{split}
\SA & = \int^{e_4}_{e_2}{e^{-3}}\,d e\int^{C_2}_{C_1}\,d C + \int^{e_3}_{e_4}{e^{-3}}\,d e\int^{C_2}_{C_3}\,d C, \\
& = \frac{2{\vp}^3}{3v(\vesc^2 - v^2)^2} + \frac{1}{8v(\vp + v)} - \frac{2\vesc + v}{6v(\vesc + v)^2}.
\end{split}
\end{equation}
In region B,
%
\begin{equation}
e_2 \le e_1 \le e_5 \le e_4 \le e_3.
\end{equation}
Hence,
%
\begin{equation}
\begin{split}
\SB & = \int^{e_5}_{e_1}{e^{-3}}\,d e\int^{C_4}_{C_1}\,d C + \int^{e_4}_{e_5}{e^{-3}}\,d e\int^{C_2}_{C_1}\,d C \\ 
& \ \ \ \ \ \ \ \ \ \ \ \ \ \ \  + \int^{e_3}_{e_4}{e^{-3}}\,d e\int^{C_2}_{C_3}\,d C, \\
& = \frac{3\vesc^2 - v^2}{3(\vesc - v)^2(\vesc+v)^2} - \frac{1}{4({\vp}^2 - v^2)}.
\end{split}
\end{equation}
In region C,
%
\begin{equation}
e_2 \le e_1 \le e_4 \le e_5 \le e_3.
\end{equation}
Hence,
%
\begin{equation}
\begin{split}
\SC & = \int^{e_4}_{e_1}{e^{-3}}\,d e\int^{C_4}_{C_1}\,d C + \int^{e_5}_{e_4}{e^{-3}}\,d e\int^{C_4}_{C_3}\,d C \\ & \ \ \ \ \ \ \ \ \ \ \ \ \ + \int^{e_3}_{e_5}{e^{-3}}\,d e\int^{C_2}_{C_3}\,d C, \\
& = \frac{3\vesc^2 - v^2}{3(\vesc - v)^2(\vesc+v)^2} - \frac{1}{4({\vp}^2 - v^2)},\\
&= \SB.
\end{split}
\end{equation}
In region D,
%
\begin{equation}
e_5 \le e_3 \le e_1 \le e_4 \le e_2.
\end{equation}
Here we can not simultaneously satisfy $e > e_1, e > e_2,$ and $e < e_3$, thus region D is forbidden. In region E,
%
\begin{equation}
e_5 \le e_1 \le e_3 \le e_4 \le e_2.
\end{equation}
So region E is forbidden for the same reason as D. Scattering events in the forbidden regions do not lead to ejection because the required kick velocity is too large. In the reference frame of the star, the DM particle must take a hyperbolic trajectory. There is therefore a limit to how much the star can `bend' the trajectory of the DM particle. Relative to the cluster, this translates to a limit on the speed that can be gained by the DM particle. Mathematically this is manifested in equations (\ref{lim3}) and (\ref{lim4}), which involve both the magnitude and direction of the kick velocity. These equations set the bounds on the integral of $\cos\delta$. In regions D and E the required lower limit on $\cos\delta$ becomes greater than the required upper limit, so this portion of parameter space is forbidden. Now Equation (\ref{esc1}) becomes,

%
\begin{equation}
\label{esc2}
\begin{split}
\bigg|\frac{1}{\Nx}\frac{\partial \Nx}{\partial t}\bigg| &= 256\pi^4\G^2\int^\infty_0{r^2}\,d r\int^\infty_0{\mpr}^2\,d \mpr \times \Bigg\{ ...  \Bigg\}, \\ 
\Bigg\{ ... \Bigg\} & = \int^{\vesc}_0{v^2f(r,v)}\,d v\int^{\vp_3}_{\vp_2}{\vp \SA g(r,\vp,\mpr)}\,d \vp \\ & \ \ \ \ \ + \int^{\vesc/3}_0{v^2f(r,v)}\,d v\int^{\vp_4}_{\vp_3}{\vp \SB g(r,\vp,\mpr)}\,d \vp \\
& \ \ \ \ \ + \int^{\vesc}_{\vesc/3}{v^2f(r,v)}\,d v\int^{\vesc}_{\vp_3}{\vp \SB g(r,\vp,\mpr)}\,d \vp  \\ & \ \ \ \ \ +\int^{\vesc/3}_0{v^2f(r,v)}\,d v\int^{\vesc}_{\vp_4}{\vp \SB g(r,\vp,\mpr)}\,d \vp.
\end{split}
\end{equation}

With the integrals over the kick velocity and given stellar mass function, it still remains to specify the stellar and DM distribution functions. We take for the stellar component a Plummer model
%
\begin{equation}
\rho_*(r) = \frac{3M_*}{4\pi}\frac{r_0^2}{(r^2+r_0^2)^{5/2}},
\end{equation}
where $r_0$ is the scale radius of the GC and the half-mass radius is $r_\mathrm h \sim 1.3r_0$ by definition. As there is little guidance on what the distribution function of DM in a GC might be, we will also use a Plummer model for the DM
%
\begin{equation} 
\rho_\chi(r) = \frac{3\MDM}{4\pi}\frac{\rx^2}{(r^2+\rx^2)^{5/2}},
\end{equation}
where $\rx$ is the scale radius of the DM halo. We choose the Plummer model for the DM in part because it has some nice mathematical properties that make it a convenient choice. Moreover, the Plummer model is reasonably realistic for GCs \cite{HenonB} and is similar to the structures of simulated DM halos and elliptical galaxies.  One shortcoming of the Plummer model is that it lacks mass segregation which is known to occur (e.g. \cite{Aarseth1966}).  This in turn implies that velocities are uncorrelated, but the error is small and there is no known analytical cluster model with mass segregation \cite{HenonB}.

\begin{figure}
\centering
\includegraphics[width=8.25cm, height=8.25cm]{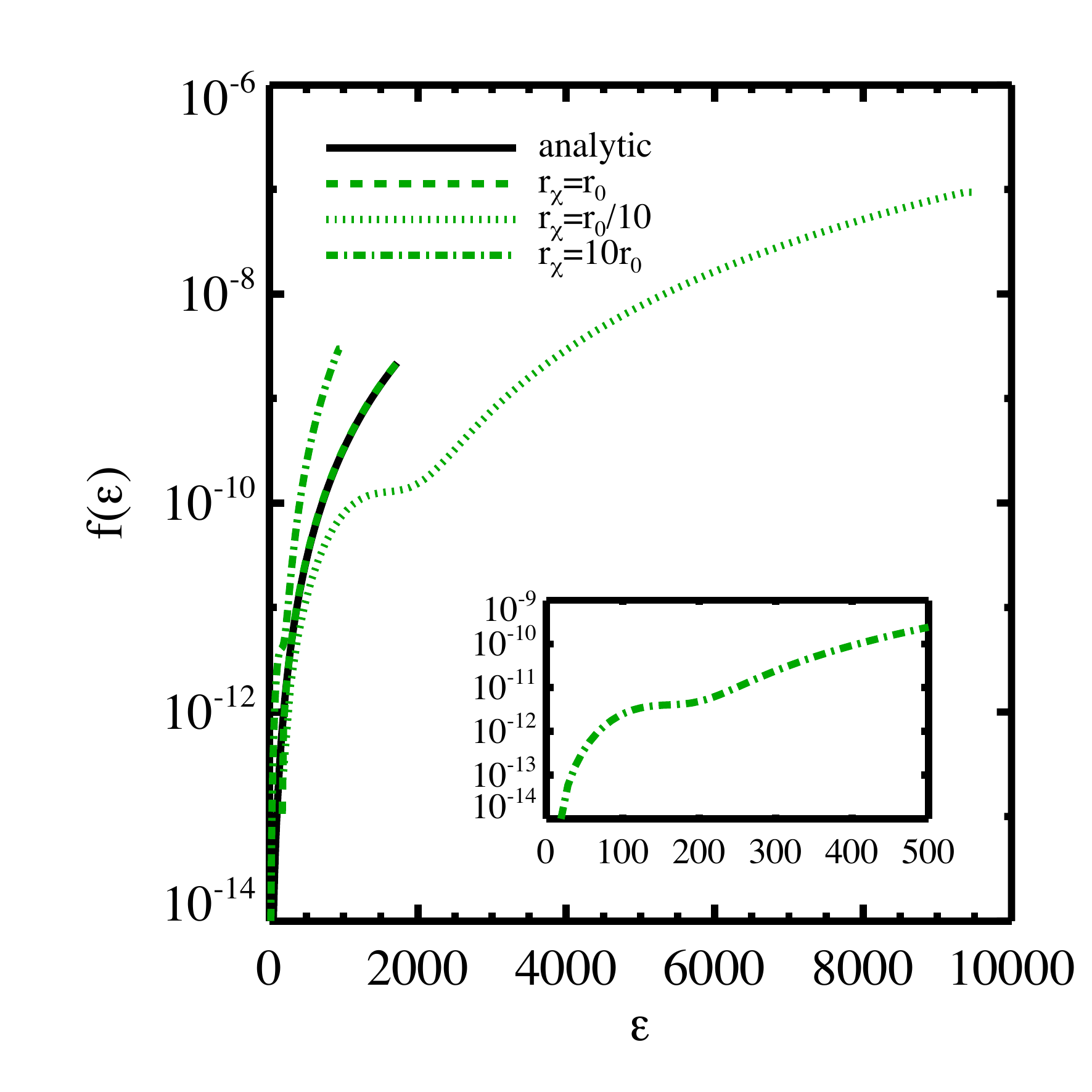}
\caption{The distribution function $f(\varepsilon)$ as a function of the magnitude of the specific energy $\big(\varepsilon = \frac{1}{2}(\vesc^2-v^2)\big)$ in the case that $M_*=2\times10^6\, \Msun,\ r_0=10\, \pc,$ and $M_\mathrm{DM} = M_*$}. The solid line is the standard Plummer model in the case that $\rx = r_0$. The dashed green line shows the numerical result for this case which is in agreement with the analytic case. The dotted line shows the distribution function in the case that $\rx = r_0/10$, while the dot-dashed line shows the case where $\rx = 10r_0$. The inset is a zoom in of the latter case, which shows the feature at $\varepsilon \approx 150$.
\label{dist func}
\end{figure}

Now the gravitational potential is
%
\begin{equation} 
\phi(r) = \frac{-GM_*}{\Big(r^2+r_0^2\Big)^{1/2}} + \frac{-G\MDM}{\Big(r^2+\rx^2\Big)^{1/2}}.
\end{equation}
In general the scale radii of the 2 components need not be the same. If $\rx \neq r_0$ the analytic expressions needed to derive the distribution function become cumbersome and we treat this case numerically. Due to the assumption of isotropy, the distribution function depends only on the magnitude of the velocity, or equivalently the kinetic energy.  Figure~\ref{dist func} shows the distribution function $f(\varepsilon)$ as a function of the magnitude of the specific energy \big($\varepsilon = \frac{1}{2}[\vesc^2-v^2]\big)$ for a GC with $M_*=2\times10^6\, \Msun,\ r_0=10\, \pc,$ and $M_\mathrm{DM} = M_*$. The solid line is the standard Plummer model in the case that $\rx = r_0$. The dashed green line shows the numerical result for this case, which is in agreement with the analytic case. The dotted line shows the distribution function in the case that $\rx = r_0/10$, while the dot-dashed line shows the case where $\rx = 10r_0$. The inset is a zoom in of the latter case, which shows the feature at $\varepsilon \approx 150\, (\km/\s)^2$.  Since $\varepsilon$ is inversely proportional to $r$, when $\rx = r_0/10$ we expect that most of the DM should be at large $\varepsilon$ (small $r$).  The flat part of the distribution function near $\varepsilon = 2000\, (\km/\s)^2$ is the transition from mostly stars at large $r$ to stars and DM at $r \sim r_0/10$.  The distribution function is also pushed to higher energies as more mass is concentrated in the center, increasing the orbital velocities in that region.  The oppositie is true for the case $\rx = 10r_0$.

In the case that $r_0 = \rx$ we will have the standard Plummer distribution,
%
\begin{equation} 
f(r,v) = \frac{24\sqrt{2}}{7\pi^3r_0^3\psi_0^5}\bigg(\frac{\vesc^2 - {v}^2}{2}\bigg)^\frac{7}{2},
\end{equation}
where $\psi_0 = \frac{\G M}{r_0}$ with $M = M_* + M_{\mathrm{DM}}$ the total mass of the cluster and $E = \frac{-3\pi\psi_0^2r_0}{64\G}$ its energy. 
%

With the choice that $\rx = r_0$ we have that 
%
\begin{equation}
\begin{split}
\vesc &= \sqrt{2\psi}, \\
&= \frac{\big(2\psi_0\big)^{1/2}}{\bigg(1+\frac{r^2}{r_0^2}\bigg)^{1/4}},
\end{split}
\end{equation}
where we have defined $\psi(r) = -\phi(r)$. 

Utilizing a Plummer model for the stellar and DM components and defining the stellar mass spectrum $N_*(m)d m$ as the number of stars in the mass interval $m \rightarrow m+d m$, we have that $g(r,\vp,\mpr) = f(r,\vp)N_*(\mpr)$. Then Equation~(\ref{esc2}) becomes
%
\begin{equation}
\label{esc3}
\begin{split}
\bigg|\frac{1}{\Nx}\frac{\partial \Nx}{\partial t}\bigg| &= \frac{2304\G^2}{49\pi^2 r_0^6\psi_0^{10}}\int^{\Rvir}_0{r^2}\,d r\int^\infty_0{N_*(\mpr)\mpr}^2\,d \mpr \times \Bigg\{ ... \Bigg\}, \\
\Bigg\{ ... \Bigg\} & = \int^{\vesc}_0{v^2(\vesc^2 - {v}^2)^\frac{7}{2}}\,d v\int^{\vp_3}_{\vp_2}{\vp \SA{(\vesc^2-{\vp}^2})^{7/2}}\,d \vp \\ & \ \ \ \ \ + \int^{\vesc/3}_0{v^2(\vesc^2 - {v}^2)^\frac{7}{2}}\,d v\int^{\vp_4}_{\vp_3}{\vp \SB{(\vesc^2-{\vp}^2})^{7/2}}\,d \vp \\ & \ \ \ \ \ + \int^{\vesc}_{\vesc/3}{v^2(\vesc^2 - {v}^2)^\frac{7}{2}}\,d v\int^{\vesc}_{\vp_3}{\vp \SB{(\vesc^2-{\vp}^2})^{7/2}}\,d \vp \\ &\ \ \ \ \ + \int^{\vesc/3}_0{v^2(\vesc^2 - {v}^2)^\frac{7}{2}}\,d v\int^{\vesc}_{\vp_4}{\vp \SB{(\vesc^2-{\vp}^2})^{7/2}}\,d \vp.
\end{split}
\end{equation}
where the virial radius $\Rvir$ of the DM halo is chosen to be suitably large $(\sim 10r_0)$ such that the integrals in Equation~(\ref{esc3}) are all converged.

We now define new variables: 
%
\begin{equation} 
x = v/\vesc, \qquad \xp = \vp/\vesc. 
\end{equation}
Then we can remove $\vesc$ from the integrals over $v$ and $\vp$ and perform those integrals separately from the radial integral. It is proven in Appendix II of ~\cite{HenonB} that the Plummer model is the only steady state distribution for which this separation is possible. Then Equation~(\ref{esc3}) becomes
%
\begin{equation}
\label{esc4}
\begin{split}
\bigg|\frac{1}{\Nx}\frac{\partial \Nx}{\partial t}\bigg| &= \frac{2304\G^2}{49 r_0^6\psi_0^{10}}\int^{\Rvir}_0{\vesc^{17}r^2}\,d r  \int^\infty_0{N_*(\mpr)\mpr}^2\,d \mpr \times \Bigg\{ ... \Bigg\} , \\ \Bigg\{ ... \Bigg\} & = \int^1_0{x^2(1 - x^2)^\frac{7}{2}}\,d x\int^{\xp_3}_{\xp_2}{\xp \SA^\prime{(1-{\xp}^2})^{7/2}}\,d \xp \\  & \ \ \ \ \ + \int^{1/3}_0{x^2(1 - x^2)^\frac{7}{2}}\,d x\int^{\xp_4}_{\xp_3} {\xp \SB^\prime{(1-{\xp}^2})^{7/2}}\,d \xp \\ & \ \ \ \ \ + \int^1_{1/3}{x^2(1 - x^2)^\frac{7}{2}}\,d x\int^1_{x_3^\prime}{\xp \SB^\prime{(1-{\xp}^2})^{7/2}}\,d \xp \\ & \ \ \ \ \  + \int^{1/3}_0{x^2(1 - {x}^2)^\frac{7}{2}}\,d x\int^1_{\xp_4}{\xp \SB^\prime{(1-{\xp}^2})^{7/2}}\,d \xp,
\end{split}
\end{equation}
where
%
\begin{equation}
\begin{split}
\xp_2 &= \frac{1}{2}(1 - x), \\
\xp_3 &= \frac{1}{2}(1 + x), \\
\xp_4 &= \frac{1}{2}(1 + 3x),
\end{split}
\end{equation}
and the $\prime$ in $S_i^\prime$ denotes the fact that it is now a function of $x$ and $\xp$ with $\vesc$ factored out.

Let us now choose a particular stellar mass spectrum. We begin with the Initial Mass Function (IMF) from ~\cite{Kroupa2001}. All of the GGCs should have ages of order $\sim10$ Gyr, meaning that their Main Sequence (MS) turnoffs should be at approximately $1\, \Msun$. Therefore, in order to obtain a crude approximation of the present day stellar mass spectrum, we simply cut off the IMF at $1\, \Msun$. Note that this is highly conservative as stellar remnants such as Neutron Stars, White Dwarfs, and Black Holes as well as any stars still on the Giant and Horizontal Branches should contribute to the escape rate. Furthermore, higher mass stars are given more weight in the integral over mass in Equation~(\ref{esc4}). The escape rate scales linearly for small variations in the mass cut, e.g. raising the cut off mass to $1.2\, \Msun$ increases the escape rate by a factor of 1.3, while lowering the cut off mass to $0.8\, \Msun$ decreases the escape by a factor of 0.7.

Given the cumbersome nature of Equation~(\ref{esc4}), it is useful to have an approximate formula for the escape rate. Performing a fit to our results below for the case $\MDM/M_*=1$
\begin{equation}
    \log_{10}\Bigg(\frac 1 N \bigg|\frac{dN}{dt}\bigg|\Bigg) \approx -23.3 + 0.500\Bigg[\log_{10}\bigg(\frac{1}{M_*r_0^3}\bigg)+13.0\Bigg].
\end{equation}
\section{RESULTS}
\label{section:results3}

\begin{figure}
\centering
\includegraphics[width=8.25cm, height=8.25cm]{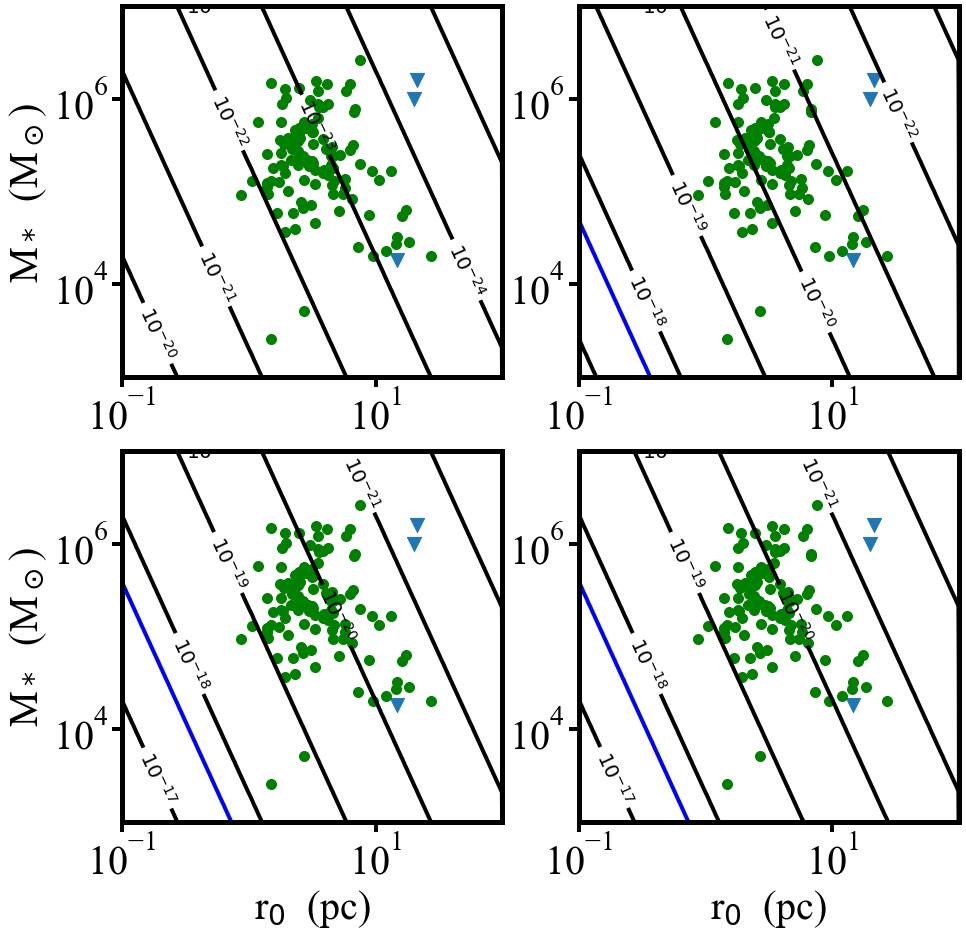}
\caption{Contours of the specific escape rate for GCs with $r_0 = \rx$. Each panel shows a different value of the ratio $\MDM/M_*$ (upper left = $10^2$, upper right = $1$, lower left = $10^{-2}$, lower right = $10^{-6}$. The blue triangles represent MGC1, an isolated cluster orbiting M31, and the isolated popluation of GGCs ($r_{gc} > 70$ kpc) that show no evidence of tidal structures. These isolated GCs are further considered in Figure~\ref{var_rad_iso}. The remaining GGCs are denoted with green circles. The solid blue line is the location where the specific escape rate is $1/\tau$ with $\tau = 13.8$ Gyr the approximate age of the Universe. GCs with escape rates comparable to or exceeding this limit should have ejected a significant portion of their DM halos. However, none of the clusters reach this limit regardless of the value of $\MDM/M_*$, with the most massive halos having the lowest escape rates as expected.}
\label{EscapePlummer1}
\end{figure}

\begin{figure}
\centering
\includegraphics[width=8.25cm, height=8.25cm]{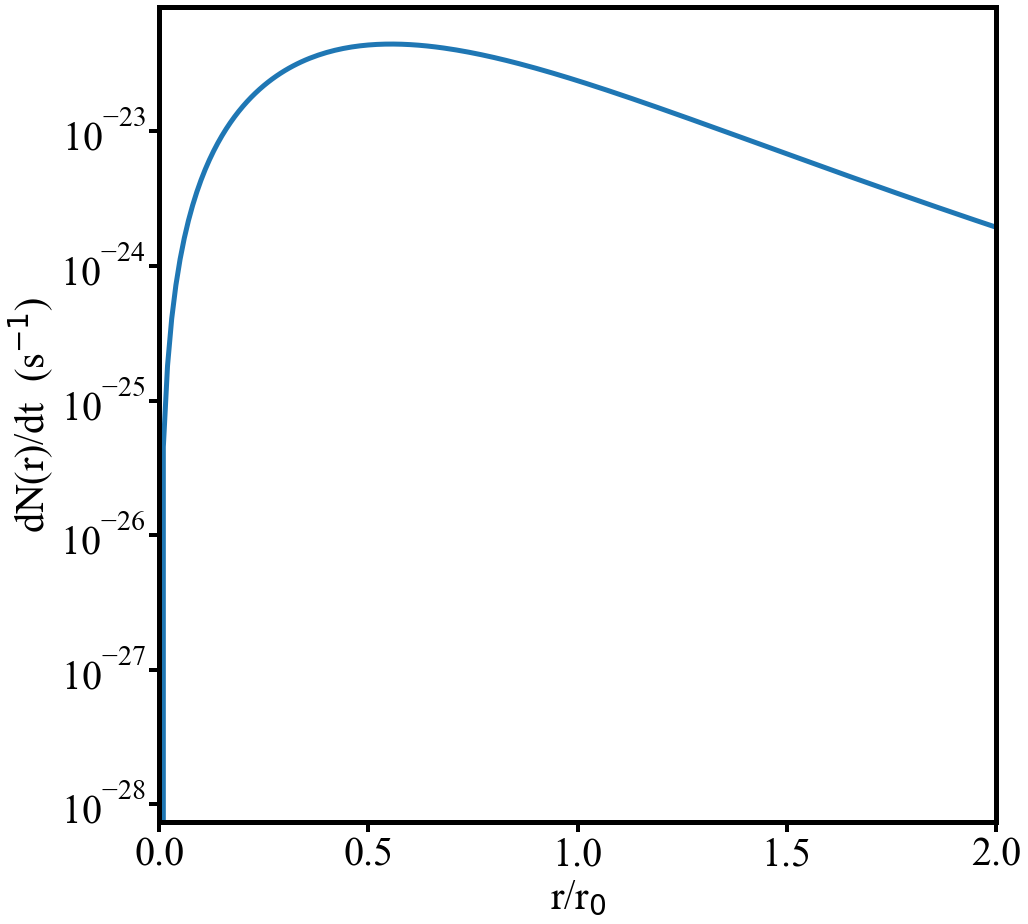}
\label{radial_rate}
\caption{The specific escape rate from a spherical shell of radius $r$ for a GC with $M_* = \MDM = 2\, \times\, 10^6\, \Msun$ and $r_0 = \rx = 10\, \pc$.}
\end{figure}

In Figure~\ref{EscapePlummer1} we consider the result of integrating Equation~(\ref{esc4}) numerically for different values of the ratio $\MDM/M_*$ and compare these results to the GGCs (as well as the cluster MGC1 located in M31). Contours of the specific escape rate for GCs with $r_0 = \rx$ are shown with solid black lines. The blue triangles represent MGC1, an isolated cluster orbiting M31, and the isolated population of GGCs ($r_{gc} > 70$ kpc) with no evidence of tidal structures. As noted in \S\ref{section:intro3}, most of the GGCs could have lost their DM halos through tidal interactions with the Galaxy. We shall therefore pay particular attention to the most isolated GCs. The green circles denote the remaining GGCs. The solid blue line is the location where the specific escape rate is $1/\tau$ with $\tau = 13.8$ Gyr the approximate age of the Universe \cite{Planck2018}. GCs with escape rates comparable to or exceeding this limit should have ejected a significant portion of their DM halos. However, none of the clusters reach this limit regardless of the value of $\MDM/M_*$.
%
Note that the escape rate is no longer sensitive to the value of $\MDM/M_*$ once this ratio has dropped below $\sim 10^{-2}$. We also note that more massive GCs have lower escape rates due to their higher escape speeds, while GCs that are larger in size have lower escape rates due to the decreased probability of experiencing an encounter at higher radii. To demonstrate this we integrate Eq.~(\ref{esc4}) with respect to all variables save for the radius in the case of a cluster with $M_* = \MDM = 2\, \times\, 10^6\, \Msun$ and $r_0 = \rx = 10\, \pc$. The resulting ``radial escape rate" is the specific escape rate from a spherical shell of radius $r$, which we denote $\frac{dN(r)}{dt}$. The radial escape rate is shown in Fig.~\ref{radial_rate}, where we see that it peaks around $r \sim 0.5r_0$ and has fallen by nearly two orders of magnitude by $r \sim 2r_0$.
%
%
%

%
\begin{figure}
\centering
\includegraphics[width=8.25cm, height=8.25cm]{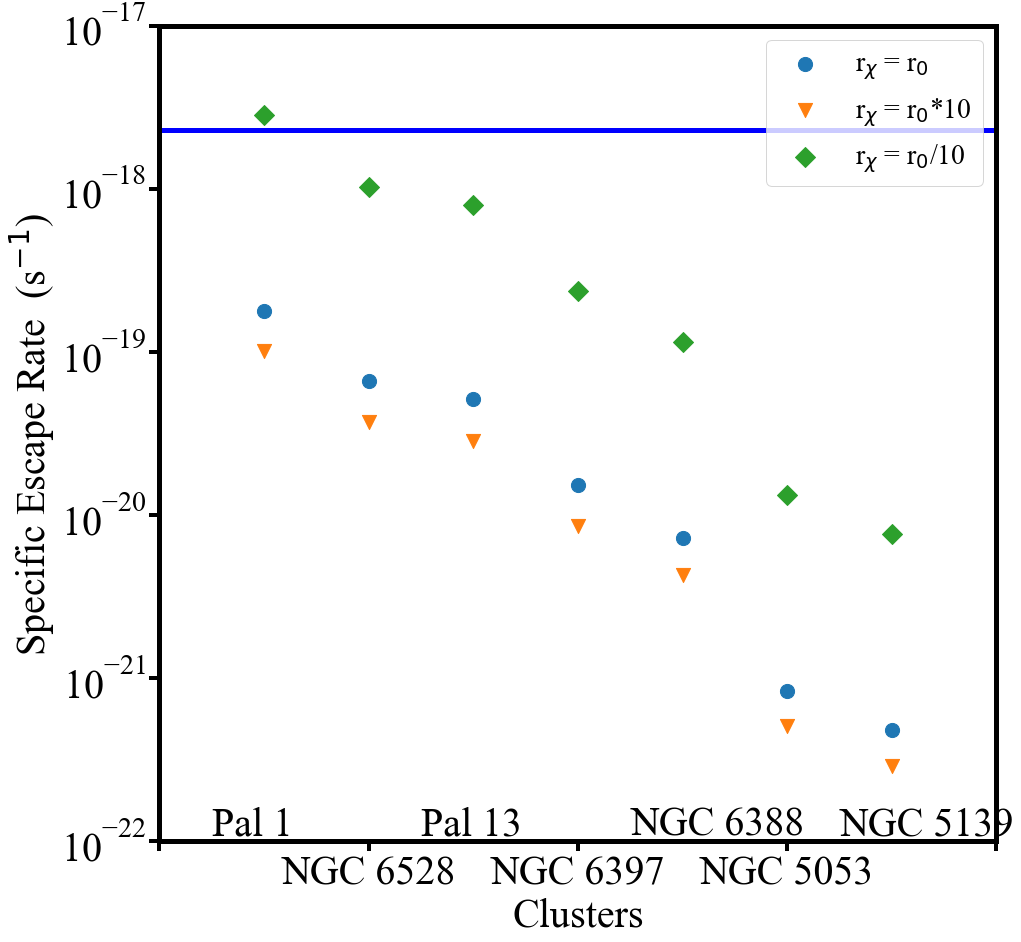}
\caption{Escape rates for several GGCs for different values of $\rx/r_0$ under the assumption that $\MDM/M_* = 1$. The solid blue line denotes $1/\tau$.}
\label{var_rad_not_iso}
\end{figure}

\begin{figure}
\centering
\includegraphics[width=8.25cm, height=8.25cm]{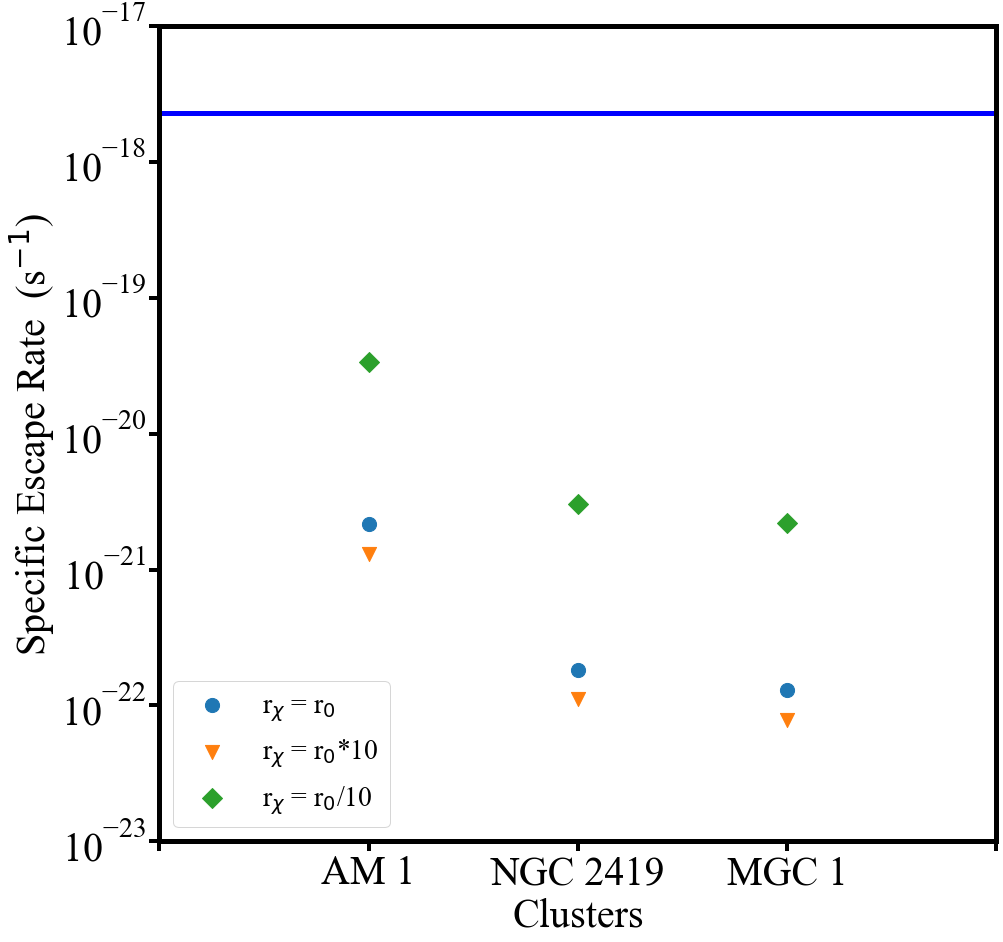}
\caption{Escape rates for the isolated GCs for different values of $\rx/r_0$ under the assumption that $\MDM/M_* = 1$. The solid blue line denotes $1/\tau$.}
\label{var_rad_iso}
\end{figure}

In Figs.~\ref{var_rad_not_iso}~\&~\ref{var_rad_iso} we consider the effect of varying $\rx$ with respect to $r_0$ while holding $\MDM/M_* = 1$. We consider $\rx~=~10r_0$, which might correspond to an extended primordial halo, as well as $\rx~=~10^{-1}r_0$, which might correspond to a cluster that has had the outer part of its DM halo stripped by tidal interactions. The results are obtained by integrating Equation~(\ref{esc2}) with the appropriate numerically derived distribution functions (see Figure~\ref{dist func}). Note that for $\rx = 10r_0$, much of the DM exists beyond the stellar content of the GC and therefore never experiences a close encounter with a star.  Thus we might normalize Equation~(\ref{esc2}) by the number of DM particles within the stellar content (say within $r_0$), rather than the total number of particles. This choice of normalization would reflect the escape rate of particles that actually experience close encounters with stars. In the case of an extended halo, slightly less than 20\% of the DM particles are within $r_0$, therefore we could multiply the results for $\rx=10r_0$ by a factor of $\sim 5$ in Figures~\ref{var_rad_not_iso}~\&~\ref{var_rad_iso}. This would still leave the escape rates from extended halos well below the figure of merit, $\tau^{-1}$.

\begin{figure}
    \centering
    \includegraphics[width = 8.25cm, height=8.25cm]{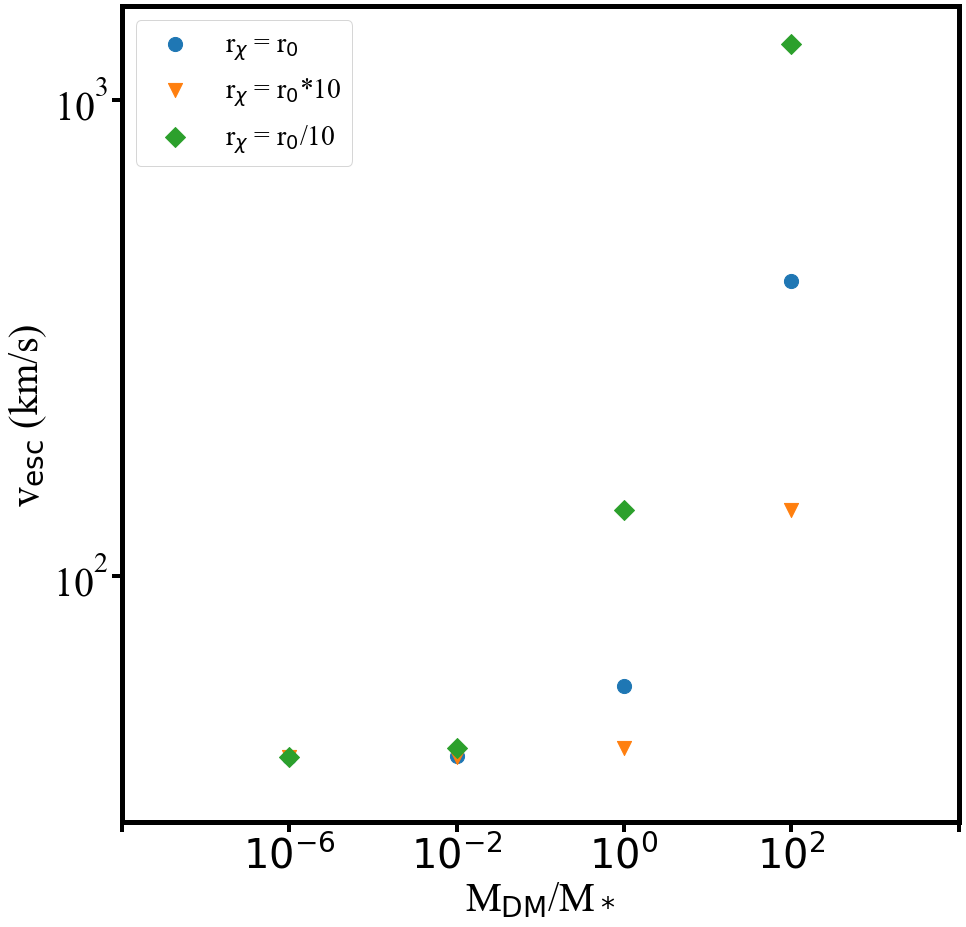}
    \caption{The central escape velocity from a GC with $M_* = 2\, \times\, 10^6\, \Msun$ and $r_0 = 10\, \pc$, for different ratios of $\MDM/M_*$ and $r_\chi/r_0$.}
    \label{escape_speeds}
\end{figure}

\begin{figure}
    \centering
    \includegraphics[width=8.25cm, height=8.25cm]{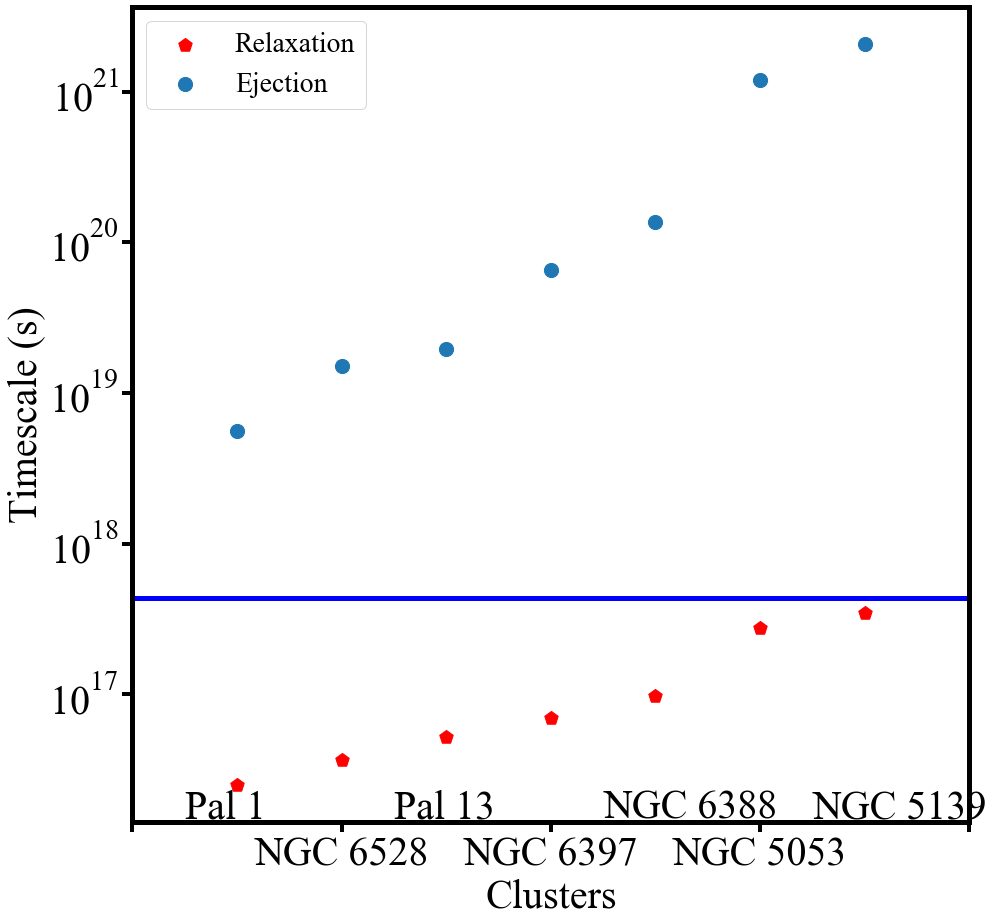}
    \caption{The relaxation and ejection time scales vary similarly with GC properties.}
    \label{relaxation}
\end{figure}

We first consider, in Figure~\ref{var_rad_not_iso}, a subset of the GGCs whose properties span those of the full distribution. The parameters for these clusters are summarized in Table~\ref{table_not_iso}, where the masses and radii are taken from \cite{Harris}, while the estimates for $M_\mathrm{DM}$ are from \cite{Shin}, and the V-band mass-to-light ratios are from \cite{Baumgardt2018}. In actuality, the observations provide projected half-light radii $R_h$. We convert these to 3D half-mass radii $r_h$ under the assumptions that mass follows light and $r_h/R_h=1.305$ as in \cite{Kowalczyk2014,Wolf2010}. Note that decreasing $\rx$ increases the escape rate. This result is perhaps counter intuitive as a smaller halo should have a deeper potential well, which is correspondingly more difficult to escape from. This can be seen in the central escape velocities, which we have plotted in  Fig.~(\ref{escape_speeds}). However, in a smaller halo, the probability of experiencing an encounter is much higher, which explains the results. Of course, the opposite is true for larger halos. Though they are easier to escape from, the probability of encounter is decreased. Note, that for $\MDM/M_* = 1$ the only halo which exceeds $1/\tau$ is that of Pal 1 in the case that $\rx = 10^{-1}r_0$. However, the escape rate can be increased by an additional half dex for smaller values of the ratio $\MDM/M_*$. Fig~\ref{var_rad_not_iso} then indicates that clusters with $\MDM/M_*~\lesssim~10^{-2}$ and $r_0$~not more than a few parsecs, could have ejected a small remnant halo after the initial halo was tidally stripped.  This also suggests that such clusters could have significantly dispersed the inner regions of their halos, even if their halos were larger ---consistent with the findings of \cite{Baumgardt2008} based on $N$-body simulations. The ejection of DM is also closely related to the two-body relaxation process, so we might expect the ejection rate to scale with cluster properties in the same way as the relaxation time. This is demonstrated in Fig~\ref{relaxation} where we have plotted the median relaxation time \cite{Harris} and the inverse of the escape rate.
%
%
%
%
%
%
\begin{table}
\centering
\begin{tabular}{ | c | c | c | c | c |}
\hline
GC & $M_* (\mathrm{M}_\odot)$ & $r_0$ (pc) & $M/L_\mathrm V$ & $\MDM/M_*$ \\
\hline
Pal 1 & $2.54\times10^3$ & 1.49 & --- & ---\\
Pal 13 & $5.12\times10^3$ & 2.72 & 10.74 & ---\\
NGC 5053 & $1.66\times10^5$ & 13.2 & 1.66 & ---\\
NGC 5139 & $2.64\times10^6$ & 7.56 & 2.90 & ---\\
NGC 6388 & $1.50\times10^6$ & 1.50 & 1.93 & ---\\
NGC 6397 & $1.59\times10^5$ & 1.94 & 2.18 & $\lesssim 1$\ \\
NGC 6528 & $9.31\times10^4$ & 0.87 & 2.26 & ---\\
\hline
\end{tabular}
\caption{Parameters for the GCs in Figure~\ref{var_rad_not_iso}.}
\label{table_not_iso}
\end{table}
%
\begin{table}
\centering
\begin{tabular}{ | c | c | c | c | c | c |}
\hline
GC & $M_* (\mathrm{M}_\odot)$ & $r_0$ (pc) & $r_{gc}$ (kpc) & $M/L_\mathrm V$ & $\MDM/M_*$\\
\hline
AM 1 & $1.81\times10^4$ & 14.7 & 124.6 & --- & ---\\
NGC 2419 & $1.60\times10^6$ & 21.4& 89.9 & 2.23 & $\lesssim 1$\ \\
MGC 1 & $1\times10^6$ & 20 & 200 & --- & $\lesssim 1$\  \\
\hline
\end{tabular}
\caption{Parameters for the isolated GCs in Figure~\ref{var_rad_iso}.}
\label{table_iso}
\end{table}

In Figure~\ref{var_rad_iso} we consider the escape rates for the most isolated clusters in the Milky Way and M31, which are marked with blue triangles in Figure~\ref{EscapePlummer1}. The parameters for these clusters are summarized in Table~\ref{table_iso}, where the masses and radii are taken from \cite{Harris}, while the limits on $\MDM$ are from \cite{Conroy, Ibata2013}, and the V-band mass-to-light ratios are from \cite{Baumgardt2018}. We convert the observed projected half-light radii to 3D half-mass radii under the assumptions discussed above. Due to their large sizes ($r_0 > 10\, \pc$), these clusters all have escape rates far below $1/\tau$. This is further evidence against the formation of GCs in DM halos.

\section{CONCLUSIONS}
\label{section:conclusions3}

GCs are peculiar systems in that they are the largest structures in the Universe not dominated by DM. Though they do not possess halos today, it is possible that they did in the past. One viable mechanism by which GCs can lose DM halos is through tidal interactions with the Galaxy. However, there exists a population of isolated GCs which should not have had their halos tidally stripped (if they ever possessed them). Observations of 2 of these GCs (NGC 2419 \& MGC1) indicate that they do not possess significant halos today ($\MDM \lesssim M_*$, see Table~\ref{table_iso}).

In this paper we have investigated an additional mechanism for the removal of DM from a GC: the ejection of DM in a single close encounter with a star. We have found that GCs could not have ejected a significant DM halo---with one exception. GCs that are sufficiently small could have ejected a small remnant halo after the majority of the halo was tidally stripped. Our results cast further doubt on the formation of GCs in extended, massive DM halos.

In the context of WIMP astronomy, GCs remain interesting targets.  As the stellar density of a GC is extremely high ($10^4-10^6\, \mathrm{stars}/\pc^3$), even a subdominant DM halo could have a density several orders of magnitude greater than that of the Solar neighborhood $\rhox \sim 0.4\, \Gev/\cm^3$.  Current limits on the mass of any hypothetical DM halo are of order the stellar mass of the cluster.  Our results indicate that such a halo could persist to the present day.

\section{Acknowledgments}
We would like to thank the referee for many insightful comments that greatly improved our manuscript. This work was funded in part by the College of Science \& Mathematics at Colorado State University - Pueblo.

\bibliographystyle{mnras}
\bibliography{Biblio}

\end{document}